\documentclass[pre,12pt]{revtex4}
\usepackage{amstext,amsfonts,amsmath,amssymb,amsthm}
\usepackage[all, knot]{xy}
\usepackage{color}
\usepackage{epsf}
\usepackage{graphics}

\newfont{\gl}{eufm10 scaled \magstep1} %% gothic fonts
\newcommand{\dd}{\hbox{d}}

\def\ee{{\rm e}}
\def\falcolor#1{#1}
\newcommand{\beq}{\begin{equation}}
\newcommand{\eeq}{\end{equation}}

\newcommand{\beqs}{\begin{equation*}}
\newcommand{\eeqs}{\end{equation*}}

\begin{document}
%\newcommand{\KeyssAndCodesMath}[2]{%
%  \begin{quote} \small
%   \textit{Keywords Mathematics:} #1.% 
%   \def\@temp{#2}
%   \ifx\@temp\@empty\relax  % <--- 3/11/06
%   \else \par\textit{MSC2000:} #2.\fi
%  \end{quote}} 
%\newcommand{\KeyssAndCodesPhys}[2]{%
%  \begin{quote} \small
%   \textit{Keywords Physics:} #1.% 
%   \def\@temp{#2}
%   \ifx\@temp\@empty\relax  % <--- 3/11/06
%   \else \par\textit{PACS numbers:} #2.\fi
%  \end{quote}} 

% ====================================================================
% Title, authors, abstract, keywords and AMS codes
% ====================================================================

\title{Extreme value distributions
 and Renormalization Group}

\author{Iv\'an Calvo}
\email{ivan.calvo@ciemat.es} 
\affiliation{ Laboratorio Nacional de Fusi\'on, Asociaci\'on EURATOM-CIEMAT,
28040 Madrid, Spain}
 \author{Juan C. Cuch\'{\i}}
 \email{cuchi@eagrof.udl.cat} 
 \affiliation{Departament d'Enginyeria Agroforestal, Universitat de Lleida,  25198
  Lleida, Spain}
 \author{J. G. Esteve\footnote{Corresponding author.}}
 \email{esteve@unizar.es}
 \affiliation{Departamento de F\'{\i}sica Te\'orica, Universidad de Zaragoza,
50009 Zaragoza, Spain}
\affiliation{Instituto de Biocomputaci\'on y F\'{\i}sica de Sistemas
Complejos (BIFI), 50009 Zaragoza, Spain}
  \author{Fernando Falceto }
%\footnote{Corresponding author.}
\email{ falceto@unizar.es}
 \affiliation{Departamento de F\'{\i}sica Te\'orica, Universidad de Zaragoza,
50009 Zaragoza, Spain}
\affiliation{Instituto de Biocomputaci\'on y F\'{\i}sica de Sistemas
Complejos (BIFI), 50009 Zaragoza, Spain}

\date{\today}

\begin{abstract}

\noindent

In the classical theorems of extreme value theory the limits of
suitably rescaled maxima of sequences of independent, identically
distributed random variables are studied. The vast
  majority of the literature on
  the subject deals with affine normalization.  We argue that more
general normalizations are natural from a mathematical and
  physical point of view and work them out.  The
problem is approached using the language of Renormalization Group
transformations in the space of probability densities. The limit
distributions are fixed points of the transformation and the study of
its differential around them allows a local analysis
of the domains of attraction and the computation of finite-size
corrections.

\end{abstract}

%\KeyssAndCodesMath{Extreme Value
% Theory}{60G70}
%\KeyssAndCodesPhys{Renormalization Group Methods}{05.10.Cc}

\maketitle

\section{Introduction}

The basic problem of Extreme Value Theory (EVT) is the following (see
Ref.~\cite{HannFerreira} for a primer). Given a 
sequence of $n$
independent identically distributed (i.i.d.) random variables, we ask
how the properly rescaled maxima of the sequence are distributed when
$n\to\infty$. Not surprisingly, EVT has much importance from the point
of view of applications in the natural
sciences~\cite{Katz,Storch,Gutenberg,Bouchaud},
finance~\cite{Embrecht}, and engineering~\cite{Weibull}, to name a
few. In all these fields one often encounters problems possessing a
threshold value for some quantity and wants to know the probability
that it be exceeded (catastrophic events are a good illustrative
example). This question is similar in spirit to that answered by the
central limit theorem, which deals with the limits of rescaled sums of
i.i.d. centered random variables. In both cases one tries to find out
whether some kind of universality exists, so that the family of limit
distributions is small and their domains of attraction are easy to
describe.

The problems of EVT and the central limit theorem are naturally
addressed in the framework of the Renormalization Group (RG), the
deepest formalism used in modern Physics to understand how a system
behaves under a change of the scale of observation. For a treatment of
the central limit theorem results and stable distributions in this
setup see Refs.~\cite{Jona1,Falceto}. Only recently has EVT been
tackled from the perspective of the Renormalization
Group~\cite{GMOR,GMORD,BG}. In the latter references, the main
motivation was to advance the understanding of the convergence to the
limit when the size of the sample data increases. Herein, we employ
the RG language to try to discuss and solve a different fundamental
problem on the acceptable rescalings and limits of maxima of sequences
of i.i.d. random variables. We describe it next.

Let $\rho$ be a probability density in ${\mathbb R}$ and $\mu$ its
distribution function,
$$\mu(x):=\int_{-\infty}^x\rho(u){\dd}u.$$
Then, the distribution function for the maximum value of $n$ i.i.d. random
variables with probability density $\rho$ is given by
$$M_n(x)=\mu(x)^n,$$
and the corresponding probability density reads
$$P_n(x)=\frac{\dd}{\dd x}M_n(x).$$ In the limit of large $n$, $P_n$
concentrates around the maximum of the support of $\rho$.  It is not
surprising that in order to obtain a non-trivial limit we have to
rescale the random variable. 
{
Since this problem was stated for the first time, 
the} {most thoroughly studied rescaling
has been} the affine one: starting from
Fr\'echet \cite{Frechet} and Fisher and Tippet \cite{FisherTippet}
there has been an extensive literature considering the possible limits
of $P_n(a_nx+b_n)$ and the domains of attraction.  Actually, Fr\'echet
only considered the case $b_n=0$ while Fisher and Tippet gave the
expression for the possible limit distributions with full generality.
Finally, Gnedenko \cite{Gnedenko} completed the solution of the problem
by describing rigorously the domains of attraction of the different
limit distributions.  But a natural question is: why to admit only
affine rescalings? At this point, nothing better than quoting from
\cite{Smith}  {(see also \cite{pancheva10})}:
\vskip 2mm

\centerline{\vbox{\hsize 12cm
\it An interesting side issue is why
this formulation was adopted at all with its affine normalization of
$M_n$. Fisher and Tippet did not explain this, whereas Gnedenko offered
only the analogy with stable distribution theory for sums, which seems 
to be begging the question. Perhaps the real explanation is that no one 
came with an alternative formulation that lead to interesting results.
The same explanation is still valid today.
}}
 \vskip 2mm 

 \noindent In this work we try to fill this gap and 
 {{motivate} the study of more general rescalings
   beyond the affine one.}  {{Actually}, this
   is not the first occasion in which non-affine rescaling is
   considered. In refs. {\cite{pancheva84,mohrav} }
   (see. also \cite{pancheva10} for a recent survey on non-linear
   rescaling) the authors explore the limit distributions and
   {domains} of attraction under power
   normalization. In \cite{mohrav} it is shown that with this
   normalization the domain of attraction of the different limit
   distributions is enlarged with respect to that obtained with affine
   rescaling. This is, in fact, {their} main
   motivation to introduce non-linear normalization.}
\falcolor{
In this work we find new reasons,
both from mathematics and from physics, to consider more general 
rescalings. These motivations will be described in next section.
}

The rest of the paper is organized as follows. In Section
\ref{sec:RGtransformation} we recast the problem of EVT into the RG
formalism and show that when the restriction of affine rescalings is
relaxed new interesting limit distributions (fixed points in the RG
approach) appear, apart from the Gumbel, Weibull, and Fr\'echet
families. In Section \ref{sec:limitdistributions} the domains of
attraction of the fixed points are studied.  Section
\ref{sec:finitesizecorrections} is devoted to finite-size corrections,
i.e. the modification of the limit distributions when the sample size,
$n$, is large but finite. In Section \ref{sec:numericaltests} some
examples and illustrative numerical tests are given. Finally, 
the conclusions are presented in Section \ref{sec:conclusions}.

\section{Renormalization Group transformation}
\label{sec:RGtransformation}

\falcolor{In this section, and in the framework of
     RG theory, we formulate the problem of EVT in such a way that
     both linear and non-linear normalizations emerge equally
     naturally under the fundamental requirement of preserving the
     support of the random variable. This gives new insights and
     allows a systematic study, as we show in subsequent sections. To
   our knowledge, the condition of support
     preservation has not been considered before and, therefore, we
   feel that we have to motivate it both from the point of view of
   mathematics and of its physical relevance.}

\falcolor{In mathematical terms we have the
   following situation: given a random variable with probability
   density $\rho$ and support $\Sigma$, it is clear that $P_n$, the
   probability density of the maximum of $n$ independent random
   variables distributed with $\rho$, is supported exactly on
   $\Sigma$.  {Hence,} if $P_n$ has the same support
   {as} the original {variable,} it is
   {natural} to require that the normalizing function
   {map $\Sigma$ onto itself, i.e. that it preserve}
   the support of the original distribution.  Note that this is in
   contrast to the distribution for sums of random variables where
   affine rescaling was first {introduced:} in this
   latter case the support of the distribution is not preserved in
   general.}
 
\falcolor{From the point of view of physics, one may argue that a
   linear normalization is more natural for it simply
   {corresponds to a change of scale or, equivalently,
     to a change of units of measurement}. This is true (and later on
   we shall deal with this case) in situations in which we consider
   dimensional quantities with {non-compact}
   support. However, in some physical {situations,}
   even {dimensional} quantities have compact
   support; for instance, in relativistic physics the velocity of a
   particle is limited by the speed of light.  In fact, very often
   relativistic velocities are {non-dimensionalized by
     using the speed of light and the modulus of the dimensionless
     velocity takes values in $[0,1]$}.  If we had a bunch of
   relativistic particles with a random distribution of velocities and
   we were interested in the distribution for the velocity of the
   fastest one it would not be very natural to obtain a limit
   distribution for velocities ranging form 0 to infinity.  The same
   can be said for spins, where the role of $c$ is played by
   {$\hbar$; or when studying, for disordered
     scattering media \cite{Popov}, the distribution of eigenvalues of
     a transfer matrix, dimensionless quantities between 0 and 1}.}

\falcolor{Recently EVT with distributions of compact support has been
considered to determine an upper bound for stellar masses \cite{OeyClarke}. 
The authors consider a Salpeter type probability density
with an upper and lower bound for the masses.
Then they determine the most massive star in groups 
with tens to hundreds of stars above the lower mass limit. 
>From the comparison of these empirical data 
and the statistical prediction they infer an upper bound for the 
stellar mass.
We will discuss this example at length in Section 
\ref{sec:numericaltests}.}

\falcolor {Motivated by the previous discussion we introduce
a RG transformation with a general rescaling of the
random variable}
\begin{equation}\label{rgtransform}
T_s \mu (x) := \mu(g_s(x))^n,
\end{equation}
where, for reasons that will become evident later, we use $s=\log n$
to parametrize it. 
\falcolor{The definition of $T_s$ requires the
choice of a rescaling function $g_s$. In the next paragraphs we will
discuss some properties to be imposed to this function.} 

In terms of the probability density $\rho$ the transformation
(\ref{rgtransform}) reads
$$T_s\rho(x)=g'_s(x)n\mu(g_s(x))^{n-1}\rho(g_s(x)).$$ Considering that
the support of $P_n$ is equal to the support of $\rho$ it is natural
to ask that $g_s$ be a homeomorphism from the support of $\rho$ onto
itself, in this way $T_s$ preserves the support of the probability
density we start with.  We remark that this condition is not imposed
in the works focusing on the affine rescaling.

The transformation $T_s$ can be extended to continuous $s$ once the
appropriate $g_s$ is defined. A natural requirement for the
transformation $T_s$ is that it forms a uniparametric group, i.e.
$$T_{s_1}\circ T_{s_2}=T_{s_1+s_2}.$$ 
Given the choice of parametrization, this holds provided that
\begin{equation}\label{grouplaw}
g_{s_2}\circ g_{s_1}=g_{s_1+s_2},
\end{equation} 
and from now on we assume that this is the case.
If we take $g_s$ differentiable with respect to $s$,
condition (\ref{grouplaw}) is equivalent to saying
that $g_s$ is solution of the differential equation
$$\frac\dd{\dd s} g_s (x) = f(g_s(x)),$$
with initial condition $g_0(x)=x$ and 
$$f(x)=\frac\dd{\dd s} g_s (x)\Big\vert_{s=0}.$$

We are actually interested in the possible extreme limiting 
distributions, i.e. in $M=\lim_{s\to\infty}T_s \mu$. This,
together with the continuity of $T_s$, 
implies that $M$ must be a fixed point of the 
{Renormalization Group} transformation,
\begin{equation}\label{fixedpoint}
M(g_s(x))^n=M(x),\ {\rm with}\ n=\ee^s.
\end{equation}
Assume that this equation has a solution with probability density
$P(x)={\rm d}M(x)/{\rm d}x$ whose support is denoted by
$\Sigma$. Several important consequences follow.
\begin{enumerate}
\item[(i)]  For $s>0$ and $x$ in the interior of $\Sigma$,  $g_s(x) > x$ 
or equivalently $f(x)>0$.
\vskip .1cm {\it Proof:} If $x$ is in the interior of $\Sigma$
the distribution function verifies
$M(x)\in(0,1)$ and is monotonically increasing. We have
$n=\ee^s>1$ and, therefore, $M(x)> M(x)^n$ and if
(\ref{fixedpoint}) holds we must have $M(g_s(x))> M(x)$, which
implies $g_s(x)> x$ and consequently $f(x)> 0$ ($f(x)=0$ implies $g_s(x)=x$).
\vskip .2cm 
\item[(ii)] If $x^*$ is at the boundary of $\Sigma$ then $f(x^*)=0$. 
\vskip .1cm 
{\it Proof:} A simple consequence of the fact that $g_s$ is a
differentiable, uniparametric group of homeomorphisms of $\Sigma$
and therefore $g_s(x^*)=x^*$.
\vskip .2cm 
\item[(iii)] $f(x^*)=0$ if and only if $x^*$ is the maximum or the
  minimum of $\Sigma$.
\vskip .1cm 
{\it Proof:} $f(x^*)=0$ implies $g_s(x^*)=x^*$ and if (\ref{fixedpoint})
holds $M(x^*)^n=M(x^*)$. But this is possible only if
$M(x^*)=0$ ($x^*$ minimum of $\Sigma$) or $M(x^*)=1$ 
($x^*$ maximum of $\Sigma$). The converse is contained in (ii).
\vskip .2cm 
\item[(iv)] $f$ can have at most two zeros and the boundary of
  $\Sigma$ at most two points. Therefore we have three possibilities:
  $\Sigma$ is the the real line $(-\infty,\infty)$, the
  semi-infinite line $[a,\infty)$ or $(-\infty,b]$, or the closed
  interval $[a,b]$.
\end{enumerate}
\vskip .2cm   

For any of the three cases mentioned in (iv) we shall take a group of
maps that preserve $\Sigma$ and study the associated RG flow.

\begin{itemize}
\item{}
{\bf Case $0$:} $\Sigma=(-\infty,\infty)$.\hfill\break
In this case a natural and simple choice for the group of
maps is the group of translations, i.e.
$$g_s(x)=x+\frac{s}\alpha, \qquad \alpha>0.$$
The most general limiting distributions (or fixed points
of the renormalization group) for this transformation is
$$M_0=\ee^{-\lambda\ee^{-\alpha x}}, \qquad \lambda>0.$$
\item{}
{\bf Case $1^-$:} $\Sigma=(-\infty,0]$.\hfill\break
The simplest choice for $g_s$ is, in this case,
$$g_s(x)=\ee^{-s/\alpha}x, \qquad \alpha>0.$$
And the corresponding limiting distribution is
$$M_1^-=\ee^{-\lambda(-x)^{\alpha}}, \qquad \lambda>0.$$
\item{}
{\bf Case $1^+$:} $\Sigma=[0,\infty)]$.\hfill\break
The maps that preserve the semi-infinite line are
$$g_s(x)=\ee^{s/\alpha}x, \qquad \alpha>0,$$
and the limiting distributions
$$M_1^+=\ee^{-\lambda x^{-\alpha}}, \qquad \lambda>0.$$
\item{}
{\bf Case $2$:} $\Sigma=[0,1]$.\hfill\break
A simple choice for the uniparametric group of maps is 
$$g_s(x)=x^{\ee^{-s/\alpha}}, \qquad \alpha>0,$$
that leads to the following family of limiting distributions
$$M_2(x)=\ee^{-\lambda(-\log x)^\alpha}, \qquad \lambda>0.$$
\end{itemize}

Note that in all cases two free positive constants $\alpha$ and
$\lambda$ appear, whose role is easy to understand: $\alpha$ fixes the
scale for the group parameter $s$ and $\lambda$ can be changed into
$\lambda\ee^s$ by the action of the group of maps, that transform a
fixed point of the RG into another one. The fixed points of Cases $0$,
$1^-$ and $1^+$ are well known in the literature and comprise the
so-called Gumbel, Weibull, and Fr\'echet distributions. While in Cases
$0$, $1^+$ and $1^-$ the rescaling is affine, it is not so in case 2.
 {The limit distributions of Case $2$ appeared in
  \cite{pancheva84} in the context of non-linear normalization. In
  {the next} section we shall study this fixed point
  and its domain of attraction under the
  {Renormalization Group} transformation.}

\section{Limit distributions with compact support}
\label{sec:limitdistributions}

Let us consider the RG transformation (\ref{rgtransform}) for
\begin{equation}\label{eq:gscase2}
g_s(x)=x^{\ee^{-s/\alpha}},
\end{equation}
 where $\alpha$ is a positive real
number. That is, we concentrate on Case 2 from Section
\ref{sec:RGtransformation}. Note in passing that the case $\alpha=1$
contains some interesting distributions among the possible fixed
points. When $\alpha=1$ the general fixed point of the transformation
is
$$M(x)=x^\lambda, \qquad\lambda>0.$$
Therefore
$$P(x)=\lambda x^{\lambda-1}$$
and, if $\lambda=1$, we get the uniform distribution.

It is easy to determine the domain of attraction of a given fixed
point when $g_s$ is of the form (\ref{eq:gscase2}). We have the
following result:

\vspace{0.5cm}

\noindent{\bf Proposition.}  A given random variable supported in
$[0,1]$ with cumulative probability distribution $\mu$ converges
weakly (or in law) after successive applications of the RG
transformation $T_s$ to $M(x)=\ee^{-\lambda(-\log x)^\alpha}$, i.e.
$$\lim_{s\to\infty}T_s\mu(x)=M(x)\quad{\rm for\ all\ } x\in[0,1],$$
if and only if 
\begin{equation}\label{limit}
\lim_{x\to1}\frac{-\log\mu(x)}{(-\log x)^\alpha}=\lambda.
\end{equation}

\vspace{0.2cm}

{\it Proof:} If (\ref{limit}) holds, then we can write
$$\mu(x)=\ee^{-\lambda(-\log x)^\alpha +o((-\log x)^\alpha)},$$
and therefore 
\begin{align}
T_s\mu(x)&=\mu(x^{n^{-1/\alpha}})^n
=\ee^{-\lambda(-\log x)^\alpha +n o\left(\frac1n(-\log x)^\alpha\right)},
\end{align}
where $n=\ee^s$. The large $s$ limit in the expression above yields
$$\lim_{s\to\infty} T_s\mu(x)=M(x)\quad {\rm for\ every\ } x\in[0,1].$$

To prove the converse note that the convergence of $T_s\mu$,
taking logarithms and for $x\not=0,1$, can be expressed as
$$\lim_{n\to\infty}n\frac{-\log\mu(x^{n^{-1/\alpha}})}{(-\log x)^\alpha}=\lambda$$
or equivalently
$$\lim_{n\to\infty}\frac{-\log\mu(x^{n^{-1/\alpha}})}{(-\log(x^{n^{-1/\alpha}}))^\alpha}
=\lambda.$$
But given that $x\not=0$ we have $\lim_{n\to\infty}x^{n^{-1/\alpha}}=1$, 
therefore $$\lim_{x\to 1}\frac{-\log\mu(x)}{(-\log x)^\alpha}
=\lambda.$$
$\,$\hfill$\Box$\break

We {emphasize that} the {appearance} of these
fixed points and attraction domains {is due} to the
non-linear rescaling function $g_s$, which in turn is motivated by the
natural requirement that the rescaling preserves the support of the
initial random variable. If we had considered the standard affine
rescaling, 
\falcolor{which could be reasonable when there is not a 
physical reason to have a bounded random variable,} 
{ the fixed points would have corresponded to the
Weibull distributions with exponent $\alpha$.}

In the next section we continue the study of the new fixed points with
the analysis of the finite-size corrections.

\section{Finite size corrections}
\label{sec:finitesizecorrections}

To discuss the amplitude of finite-size corrections and their shape,
i.e. the behavior of the extremal distributions when the number of
i.i.d random variables $n$ is large but finite, we must study the
neighborhood of the fixed points and the linear approximation of the
RG transformation (\ref{rgtransform}). For this we compute its
differential at a probability distribution $\mu$ acting on $\eta$:
\begin{equation}\label{sec:defdifferential}
(DT_s)_\mu \eta = n \mu(g_s(x))^{n-1}\eta(g_s(x)).
\end{equation}
The stable and unstable directions of a fixed point $\mu$ are given by
the eigenvalues and eigenvectors of (\ref{sec:defdifferential}) at
$\mu$.  They determine the amplitude of the finite-size corrections
and their shape.

We focus on the case $g_s(x)=x^{n^{-1/\alpha}}$, with $n=\ee^s$ and
fixed point $M(x)=\ee^{-\lambda(-\log x)^\alpha}$.  In order to solve the 
eigenvalue equation for (\ref{sec:defdifferential}) 
it is very useful to consider the following ansatz
$\eta(x)=M(x)\phi(x)$. In terms of it, the eigenvalue equation reads
$$(DT_s)_\mu M(x)\phi(x) = \nu M(x)\phi(x),$$
that due to the
properties of the fixed point $M(x)$ reduces to
$$n\phi(g_s(x))=\nu \phi(x).$$
This is solved by
$$\phi_{_\beta}(x)=(-\log x)^\beta,$$ with eigenvalue
$\nu_{_\beta}=n^{1-\beta/\alpha}$. A perturbation of the fixed point
is unstable (or relevant, in the RG terminology) if the corresponding
eigenvalue is greater than one, i.e. $\beta<\alpha$ and it is stable
(irrelevant) if $\beta>\alpha$. The case $\beta=\alpha$ consists of a
perturbation tangent to the line of fixed points and, therefore it
corresponds to a purely marginal direction.

Note that the above analysis is consistent with the domains of
attraction determined in Section \ref{sec:limitdistributions}. The
stable directions are precisely those that do not alter the limit in
(\ref{limit}), the marginal ones induce an infinitesimal change in the
limit and therefore also in the fixed point to which the perturbed
distribution tends, and finally an unstable perturbation makes the
limit diverge, implying that the perturbed distribution does not
converge under successive applications of the RG transformation.

To understand how the linear analysis above is useful to determine
the finite size corrections, consider the following situation.
We start with a random variable with cumulative distribution 
$\mu(x)$ expanded in the eigenvectors obtained above, 
\begin{equation}\label{expansion}
\mu(x)=M(x)(1+\sum_ic_i\phi_{_{\beta_i}}(x)),
\end{equation}
where the terms in the sum are ordered 
according to their eigenvalues, 
so that $\nu_{_{\beta_i}}>\nu_{_{\beta_j}}$ for $i<j$.

Assuming that all eigenvalues are smaller than one ($\beta_i>\alpha$)
or, in other words, that $\mu(x)$ belongs to the domain of attraction of $M(x)$,
 one can show that
$$
T_s\mu(x)=M(x)+
c_1 n^{1-\beta_1/\alpha}M(x)\phi_{\beta_1}+ 
         o(n^{1-\beta_1/\alpha}).
$$ Hence, the largest eigenvalue determines the behavior with $n$ (the
         size of the system) of the amplitude of the dominant
         correction while its eigenfunction determines the shape of
         the correction. One can also study corrections of higher
         order and go beyond the linear approximation. In the next
         section we show how to accomplish this and compare our
         approximations with numerical implementations of the
         statistical models to test their reliability.

\section{Examples. Numerical tests}
\label{sec:numericaltests}
 
\falcolor{
We start this section by discussing the example presented in the 
introduction that has been used in \cite{OeyClarke} to determine 
an upper mass limit to the stellar initial mass function.
To be specific consider the Salpeter probability distribution for 
massive stars
$$\sigma(m)=a_0 m^{-2.35},\quad 10<m<200,\quad a_0=30.7618...$$
where masses are expressed in solar mass units.}

\falcolor{
For a systematic treatment of the problem it is more
convenient to rescale the random variable to another
one supported in $[0,1]$. We define $x=(m-M_{\rm lo})/(M_{\rm up}-M_{\rm lo})$ 
where $M_{\rm lo}=10$ and $M_{\rm up}=200$ are the lower and upper limits
of the distribution. In terms of this variable we get the 
probability density
$$\rho(x)= a(1+19x)^{-2.35},\quad 0<x<1,\quad a=26.1075...$$}

\falcolor{
We can determine now the appropriate scaling in the
renormalization group transformation,
that corresponds to $\alpha=1$, and the corresponding fixed point
$$\lim_{n\to\infty}\rho_n(x)=\lambda x^{\lambda-1},\quad \lambda=0.0228741...$$
where $\rho_n=T_{\log n}\rho$.}

\falcolor{
The finite size corrections can be computed as well to give
$$\rho_n(x)=\lambda x^{\lambda-1}\left(1-\frac{c_2}n\log(x)
(\log(x)+\frac2\lambda)+ o(n^{-1})\right),\quad c_2=0.0143578...$$
}

In order to test our theoretical predictions we make two numerical
experiments in which we study the distribution for the maxima of $n$
independent random variables with a probability density $\rho$. The
actual size of the systems is chosen so that the perturbative approach
discussed above applies and the experiment is repeated a number of
times large enough to make the statistical error much smaller than the
finite size corrections. The numerical simulations are performed by
generating $n$ independent random variables and selecting their
maximum.  We divide the interval into 50 bins and after repeating the
experiment $N$ times we obtain the frequency with which the maximum
belongs to a given bin.  The frequency, properly normalized, will be
our numerical approximation to $\rho_n=T_s \rho$, with $n=\ee^s$.

\begin{figure}
\begin{center}
\epsfxsize=10cm
\epsffile{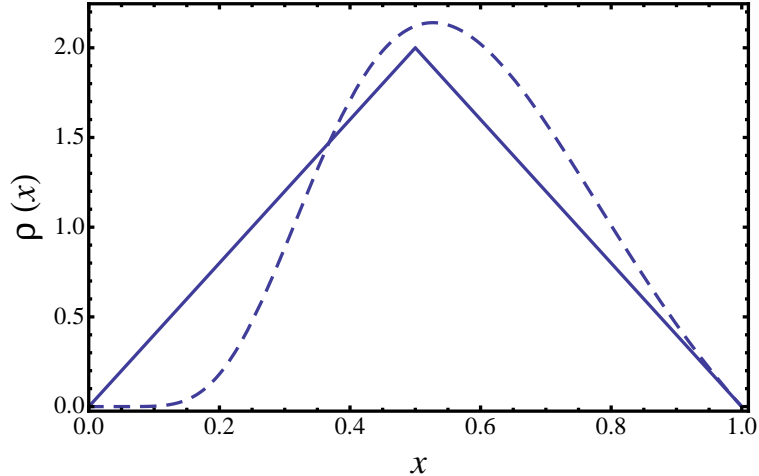}
\end{center}
\caption{Probability 
density function of (\ref{ejemplo1}) (solid line)
and its limiting function (dashed line).}
\label{fig1}
\end{figure}

Our first example for $\rho$ 
is the tent distribution, whose probability
density is given by
$$
\rho(x)=
\begin{cases}
4x, & x\le 1/2,  \\
4-4x, & x>1/2.
\end{cases}
$$
Observe that the support of $\rho$ is the interval $[0,1]$.
It is plotted, together with the density
of its limiting 
distribution, 
in Fig. \ref{fig1}

The cumulative distribution function determined by $\rho$ is
\begin{equation}\label{ejemplo1}
\mu(x)=
\begin{cases}
2x^2, & x\le 1/2,  \\
1-2(1-x)^2, & x>1/2,
\end{cases}
\end{equation}
that converges under the action of the RG transformation
for $\alpha=2$ to the cumulative distribution function
$$
M(x)=\ee^{-2(-\log x)^2}.
$$

If we perform the expansion in (\ref{expansion}) we obtain
$$
\mu(x)=M(x)(1+2(-\log x)^3-\frac{19}6(-\log x)^4+\frac92(-\log x)^5+\cdots).
$$ The most relevant (or rather the least irrelevant) eigenvalue in
the expansion is $\nu_3= n^{-1/2}$ and it determines the behavior
with $n$ of the amplitude of the finite-size corrections. In order to
quantify the corrections when the number of random variables is $n$,
we use the $L^1$ norm for the difference of the probability
densities. This norm is also called total variation metric in the
context of probability theory (see \cite{HaanRes} and references
therein).  We expand
\begin{equation}\label{finitesize}
\Delta:=\int_0^1|\rho_n(x)-M'(x)|\dd x = 
c_{1/2} n^{-1/2}+c_{1} n^{-1}+c_{3/2} n^{-3/2}+\cdots
\end{equation}
where $\rho_n=T_s \rho$ with $n=\ee^s$.
The first coefficient is given by
$$ 
c_{1/2}=2\int_0^1\left|\frac{\rm d}{{\rm d}x}[M(x)\log(x)^3]\right|\ \dd x=\frac32\sqrt3\ee^{-3/2},
$$
and, similarly, one can compute the others to obtain
$$c_1=-\frac{15}8\ee^{-3/2},\qquad c_{3/2}=\frac9{32}\sqrt3\ee^{-3/2}, \cdots$$

\begin{figure}
\epsfxsize=10cm
\begin{center}  
\leavevmode   
\epsffile{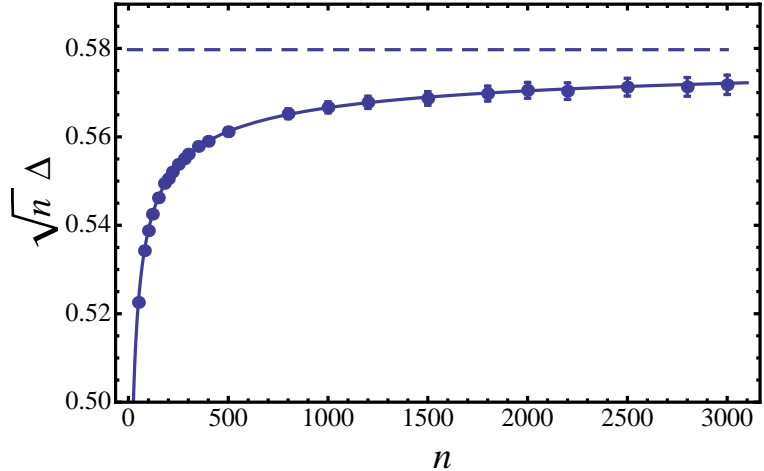}
\end{center}
\caption{Finite size corrections rescaled with $\sqrt n$ as a function
  of the size of the system $n$. Dots represent the values obtained
  in the numerical experiment, with error bars corresponding to two standard
deviations, and the solid line is the
  theoretical prediction for $\sqrt{n}\Delta$ up to the $n^{-1}$
  term.}
\label{fig2}
\end{figure}

In Fig. \ref{fig2}
we have plotted the finite-size corrections
to the distribution of the maxima obtained numerically
scaled with $\sqrt n$, for different values 
of the size of the system $n$. We observe a very good agreement with the 
theoretical predictions in (\ref{finitesize}). 

The second prediction that we test numerically is the shape of the
corrections. In this case we take a fixed (and large) value for $n$
and we plot the rescaled difference between the limiting distribution
and the one obtained numerically for the maxima of $n$ random
variables distributed according to $\rho$.  The finite-size
corrections $\delta(x):=(\rho_n(x)-M'(x))$ can be expanded as
$$\delta(x)=\delta_{1/2}(x)n^{-1/2}+\delta_1(x)n^{-1}+\cdots,$$
with the first coefficients given by
\begin{align}\label{deltas}
\delta_{1/2}(x)&=
-2\frac{\rm d}{{\rm d}x}
\big[ M(x)\log(x)^3\big],\cr
\delta_1(x)&=
\frac{\rm d}{{\rm d}x}\big[ M(x)\log(x)^4(\frac{19}6 - 2\log(x)^2)\big].
\end{align}

In Fig. 3 the dots
represent the points obtained with the numerical experiment for 
$\sqrt n\delta(x)$
corresponding to $n=3000$. The error bars, a little larger than the 
size of the dots, represent the statistical 
uncertainty due to the limited size of the sample.
The dashed line is $\delta_{1/2}(x)$ as defined in (\ref{deltas})
while the solid line includes the next correction 
$\delta_{1/2}(x)+\delta_1(x)n^{-1/2}$.  
We see an excellent agreement between the theoretical 
prediction and the numerical experiment, especially when the 
subleading correction is included.

\begin{figure}
\epsfxsize=10cm
\begin{center}
\leavevmode   
\epsffile{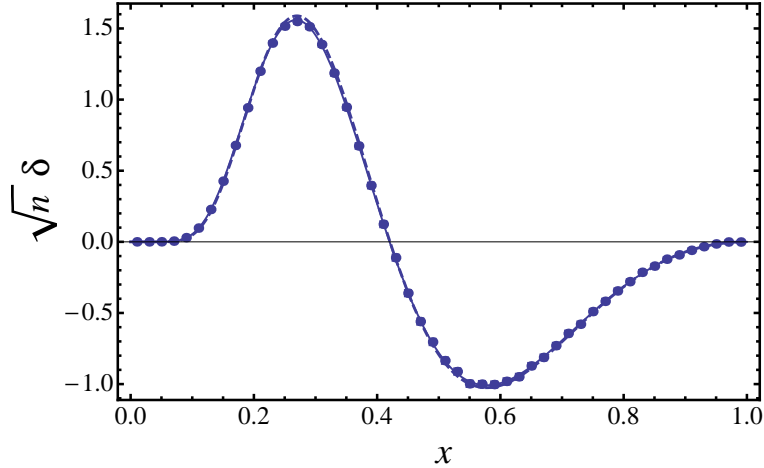}
\end{center}
\caption{Shape corrections for $n=3000$. Dots represent the outcome of
  the numerical experiment. The error bars, representing two standard
  deviations, are of the order of the dot size in the plot.  The dashed
  line is the leading term of the correction $\delta_{1/2}(x)$ while
  the solid line includes the next subleading term
  $\delta_{1/2}(x)+\delta_1(x)n^{-1/2}$.}
\label{fig3}
\end{figure}

\begin{figure}
\epsfxsize=10cm
\begin{center}
\leavevmode   
\epsffile{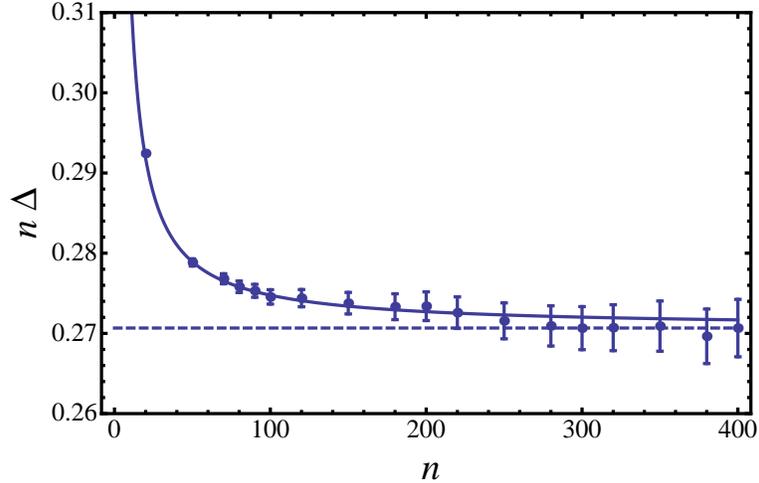}
\end{center}
\caption{Finite size corrections in the second example scaled with
  $n$.  In the $x$ axis we represent the size of the system, $n$.
  Dots are the values obtained from the numerical experiment and the
  solid line is the theoretical prediction (\ref{finitesize2}) for
  $n\Delta$ up to the $n^{-2}$ term.}
\label{fig4}
\end{figure}

The second example has a probability density
$$
\rho(x)=
|2-4x|,
$$
and a cumulative distribution function
$$
\mu(x)=
\begin{cases}
2x(1-x), & x\le 1/2,  \\
1-2x(1-x), & x>1/2.
\end{cases}
$$ It converges under the action of the RG with
$\alpha=1$ to $M(x)=x^2$. The limiting probability density is
$M'(x)=2x$. We can expand again,
$$
\mu(x)=M(x)(1 + (-\log x)^2 + (-\log x)^3 +\frac7{12}(-\log x)^4+\cdots),
$$ and we find that the most relevant perturbation has an eigenvalue
$\nu_2=n^{-1}$, which determines the leading behavior with $n$ of the
amplitude of the finite-size corrections. If we also keep the first
subleading terms we obtain
\begin{equation}\label{finitesize2}
\Delta=\int_0^1|\rho_n(x)-M'(x)|\dd x = 
c_{1} n^{-1}+c_{2} n^{-2}+c_{3} n^{-3}+\cdots
\end{equation}
with 
$$
c_1=2\ee^{-2},\quad c_2=3\ee^{-2},\quad c_3=\frac52\ee^{-2}.
$$
In Fig. \ref{fig4} we check the validity of this expansion. We see that
within the statistical errors, due to the limited size of the sample,
the finite-size corrections agree with the theoretical predictions.

The shape of the corrections in this case is
\begin{equation}\label{finiteshape2}
\delta(x)=\delta_1(x)n^{-1}+\delta_2(x)n^{-2}+\cdots
\end{equation}
with the different contributions given by
\begin{align}\label{deltas2}
\delta_{1}(x)&=
2x\log x(1-\log x),\cr
\delta_2(x)&=
x\log(x)^2(\log(x)^2-3).
\end{align}

\begin{figure}[h!]
\epsfxsize=9.89cm
\begin{center}
\leavevmode   
\epsffile{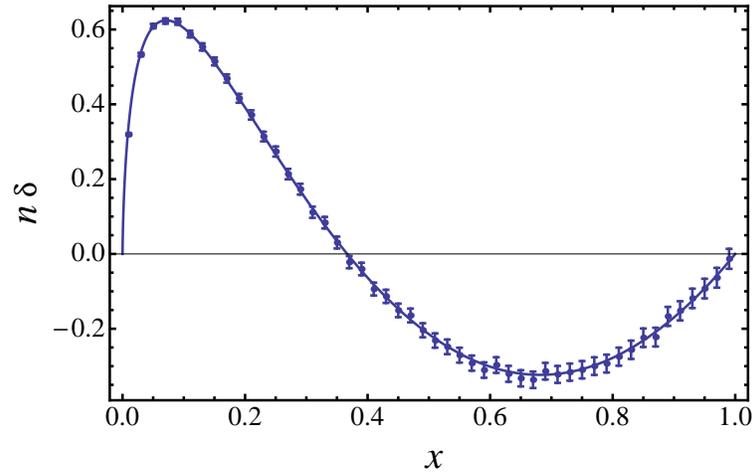}
\end{center}
\caption{Shape corrections for the second example.  Here $n=300$. Dots
  represent the outcome of the numerical experiment. The error bars
  stand for two standard errors.  The dashed line is the leading term
  of the correction, $\delta_{1}(x)$, while the solid line includes
  the next subleading term, $\delta_{1}(x)+\delta_2(x)n^{-1}$. The
  dashed line almost coincides with the solid line; only near the
  maximum the small difference can be appreciated.}
\label{fig5}
\end{figure}

In Fig. \ref{fig5} we show the numerical value for the shape
correction and compare it with the analytical prediction in
(\ref{finiteshape2}).  We can see again a remarkable agreement between
the numerical experiment and the theoretical prediction.

In the previous examples we have tested the accuracy of the finite
size analysis carried out in Section
\ref{sec:finitesizecorrections}. The size of the system and the sample
have been chosen so that the computational time is reasonable and the
errors are sufficiently small not to spoil any predictive power.
Within this range we have to go beyond first order corrections
(represented by the dashed line in Figs.~\ref{fig3} and \ref{fig5}) to
give a precise description of the experimental results. The number
of terms needed depends, of course, on the concrete
details of the problem and the required accuracy.

\section{Conclusions}
\label{sec:conclusions}

By employing Renormalization Group techniques we have studied the
limit distribution of the appropriately rescaled maximum value of a
sequence of $n$ independent, identically distributed random variables
when $n\to\infty$. {Obviously, the rescaling is needed
  for obtaining non-trivial limits. Most of the literature on Extreme
  Value Theory is devoted to the study of these limits under affine
  rescalings, perhaps by analogy to the treatment of the problem of
  stable distributions. However, when computing limits of sequences of
  maxima of independent, identically distributed random variables, it
  seems natural to impose that the rescaling preserves the support of
  the original random variable, a condition that the affine rescaling
  does not meet, in general.}

We have recast the problem of finding such limit distributions into
the language of the Renormalization Group, explained how the condition
of support preservation naturally arises, and what its implications
are. {The main contribution of this paper is showing
  that in this framework linear and non-linear rescalings are treated
  on an equal footing, and which one should be employed follows
  precisely from support preservation. In our formulation the limit
  distributions are fixed points of the Renormalization Group
  transformation. After the identification and discussion of the fixed
  points we have worked out the differential of the transformation
  around them, with emphasis on those associated to non-linear
  rescalings. This helps understand the domains of attraction and the
  corrections due to large but finite $n$, the so-called finite-size
  corrections.}

An interesting technical aspect of the approach herein adopted is the
concrete form of the definition of the Renormalization Group
transformation. We define it as an uniparametric group of
transformations that is fixed once for all, differing from other works
in this line where the transformation can be adapted at every step.
This fact has some consequences, especially concerning the domain of
attraction of the fixed points. Indeed, within our approach the
determination of the domain of attraction is simpler.
One may wonder whether it is possible to modify the classical
results on the domains of attraction when we restrict to
transformations that preserve the support of the original random
variable.

\falcolor{Finally, we would like to mention the fact that our RG fixed points 
are trivial in the sense that they correspond to EV for independent 
random variables.
It would be interesting to study how these results 
(rescaling and fixed points) are to be modified 
when considering a family of correlated
random variables.}

\vskip 1cm
\noindent{\bf Acknowledgements:} I.~C. acknowledges the hospitality of
the Department of Theoretical Physics at the University of Zaragoza,
where part of this work has been done. Research partially supported by
grants 2009-E24/2, DGIID-DGA and FPA2009-09638,
ENE2009-07247, Ministerio de Ciencia e Innovaci\'on (Spain).

%%%%%%%%%%%%%%%%%%%%%%%%%%%%%%%%%%%%%%%%%%%%%%%%%%%%%%%%%%%%%
%%%%%%%%%%%%%%%%%%%%%%%%%%%%%%%%%%%%%%%%%%%%%%%%%%%%%%%%%%%%%%%%%%


\begin{thebibliography}{99}
%\expandafter\ifx\csname natexlab\endcsname\relax\def\natexlab#1{#1}\fi
%\expandafter\ifx\csname bibnamefont\endcsname\relax
%  \def\bibnamefont#1{#1}\fi
%\expandafter\ifx\csname bibfnamefont\endcsname\relax
%  \def\bibfnamefont#1{#1}\fi
%\expandafter\ifx\csname citenamefont\endcsname\relax
%  \def\citenamefont#1{#1}\fi
%\expandafter\ifx\csname url\endcsname\relax
%  \def\url#1{\texttt{#1}}\fi
%\expandafter\ifx\csname urlprefix\endcsname\relax\def\urlprefix{URL }\fi
%\providecommand{\bibinfo}[2]{#2}
%\providecommand{\eprint}[2][]{\url{#2}}

\bibitem{HannFerreira} L. de Haan and A. Ferreira, {\it Extreme Value
  Theory: An Introduction} (Springer, 2006).

\bibitem{Katz} R. W. Katz, M. B. Parlange, and P. Naveau, Adv. Water
  Resour. {\bf 25}, 1287 (2002).
 
\bibitem{Storch} H. v. Storch and F. W. Zwiers, {\it Statistical
  Analysis in Climate Research} (Cambridge University Press,
  Cambridge, 2002).

\bibitem{Gutenberg} B. Gutenberg and C. F. Richter, Bull. Seismol. Soc. Am. 
{\bf 34}, 185 (1944). 

\bibitem{Bouchaud} J.-P. Bouchaud and M. M\'ezard, J. Phys. A {\bf
  30}, 7997 (1997).

\bibitem{Embrecht} 
P. Embrecht, C. Kl\"uppelberg, and T. Mikosch,
  {\it Modelling Extremal Events for Insurance and Finance} (Springer,
  Berlin, 1997).

\bibitem{Weibull} 
W. Weibull, J. Appl. Mech.-Trans. ASME {\bf 18}, 293 (1951).

\bibitem{Jona1}
G. Jona--Lasinio, Nuovo Cimento {\bf 26}, 98 (1975).

\bibitem{Falceto}
I. Calvo, J. C. Cuch\'{\i}, J. G. Esteve and  F. Falceto,
J. Stat. Phys. {\bf 141}, 409 (2010).

\bibitem{GMOR}
G. Gyorgyi, N. R. Moloney, K. Ozogany and Z. Racz, Phys. Rev. Lett. {\bf 100},
 210601 (2008).

\bibitem{GMORD}
G. Gyorgyi, N. R Moloney, K. Ozogany, Z. Racz and M. Droz, 
Phys. Rev. E {\bf 81}, 041135 (2010).

\bibitem{BG}
E. Bertin and G. Gyorgyi, J. Stat. Mech., P08022 (2010).

\bibitem{Frechet} M. Fr\'echet, Ann Soc. Polonaise Math. {\bf 6}, 93 (1927).

\bibitem{FisherTippet} M.R. Fisher and L.H.C. Tippet, Proc. Cambridge Philos. 
Soc. {\bf 24}, 180 (1928).

\bibitem{Gnedenko} B.~V. Gnedenko, Ann. Math {\bf 44}, 423 (1943).
 
\bibitem{Smith}
R.L. Smith, 
Breakthroughs in statistics,
(S. Kotz and N. L. Johnson eds.) Springer-Verlag,
(1993).

\bibitem{pancheva10}
E. Pancheva, ProbStat Forum {\bf 3}, 11 (2010).

\bibitem{pancheva84}
E. Pancheva, Lecture Notes in Math., {\bf 1155}, 284 (1984).


\bibitem{mohrav}
N.R. Mohan and S. Ravi, Theory Probab. Appl. {\bf 37}, 632 (1991).


\bibitem{Popov} S. M. Popov, G. Lerosey, R. Carminati, M. Fink, A. C. Boccara and S. Gigan, Phys. Rev. Lett. {\bf 104}, 100601 (2010).


\bibitem{OeyClarke}
M.S. Oey and C.J. Clarke, Astrophysical J. {\bf620}, L43-46 (2005).

\bibitem{HaanRes}
L. de Haan and S. Resnick, The Annals of Probability {\bf24}, 97 (1996).


\end{thebibliography}
\end{document}